\documentclass[11pt,a4paper,final]{article}
\usepackage[utf8]{inputenc}
\usepackage[english]{babel}
\usepackage{amsmath}
\usepackage{amsfonts}
\usepackage{amssymb}

\def\be{\begin{equation}}
\def\ee{\end{equation}}
\def\bea{\begin{eqnarray}}
\def\eea{\end{eqnarray}}

\newcommand{\nn}{\nonumber}

\newcommand\diff{\mathrm{d}}

\newcommand{\dd}{\mathrm{d}}

\newcommand{\ii}{\mathrm{i}}

\newcommand{\mt}{\mathrm{T}}

\newlength{\sswidth}

\newcommand{\hook}{\mathbin{\rule[.2ex]{.4em}{.03em}\rule[.2ex]{.03em}{.9ex}}}
\newcommand{\alphaangle}{\vartheta}

\newcommand{\Free}{\mathcal{F}}

\author{Carolina Matté Greogry}
\title{Classifying solutions to Romans supergravity with a zero B-field}

\begin{document}

\bibliographystyle{utphys}

\title{Classifying solutions to Romans supergravity \\ with a zero B-field}

\author{\\
\\
Carolina Matté Gregory\thanks{carolina.gregory@unb.br}\\
\\
\\
\emph{Instituto de Física, Universidade de Brasília}\\
\emph{70910-900, Bras\'{\i}lia, DF, Brasil}\\
\\
\\
}
\date{}
\maketitle
\vspace{2cm}
\begin{abstract}
\noindent Expanding on previous papers, we continue studying Euclidean Romans supergravity in six dimensions with a non-trivial Abelian R-symmetry gauge field. Using a set of differential constraints on a $SU(2)$ structure, we look for further geometric solutions to such equations when we turn off the two-form potential B. We find that the six-dimensional space is described by a two-dimensional fibration over a four-dimensional manifold with a Kähler metric. We then classify these types of solutions.

\end{abstract}
\vspace{5cm}
\pagebreak


\section{Introduction}

In previous papers \cite{Alday:2014bta,Alday:2015jsa}, we constructed gravity 
duals to five-dimensional gauge theories on non-conformally flat backgrounds, specifically, 
certain families of squashed five-spheres and Sasaki-Einstein manifolds. In \cite{Alday:2014bta}, we have used the six-dimensional Romans $F(4)$ supergravity 
\cite{Romans:1985tw}, which is a consistent truncation of massive IIA supergravity 
on $S^4$ \cite{Cvetic:1999un}, and, more recently, it has been shown to be also a consistent truncation of IIB supergravity on a warped product of $S^2 \times \Sigma$, where $\Sigma$ is a Riemann surface \cite{Hong:2018amk,Malek:2018zcz}.
 Having constructed these supergravity 
solutions, 
we then computed the holographic free energy $\Free=-\log Z$ by 
holographically renormalizing the on-shell Euclidean action. 
The perturbative partition function for these theories has been computed in \cite{Imamura:2012xg} and we have explicitly shown that the large $N$ limit of these partition functions is in precise agreement with the holographic free energies of  our supergravity solutions.

In \cite{Alday:2015jsa}, we have shown that real Euclidean supersymmetric solutions to 
Romans $F(4)$ gauged supergravity, with 
a non-trivial Abelian R-symmetry gauge field, have a canonical $SU(2)$ structure determined by the Killing spinor. 
More precisely, we have shown that the supersymmetry conditions together with the equations of motion
are equivalent to a set of differential constraints on this $SU(2)$ structure. 
This geometric formulation then led to a number of interesting applications. 
We showed that this  structure extended into the bulk the 
conformal boundary $SU(2)$ structure studied in \cite{Alday:2015lta}, which allowed for the construction of gravity duals 
to families of five-dimensional gauge theories on rigid backgrounds.  
As another application we extended several of the results of the previous paper. 

In fact, analysing supersymmetric solutions via G-structures has been widely used, first shown in \cite{Gauntlett:2002sc}, then developed for various string theory settings    \cite{Gurrieri:2002wz}-\cite{Gauntlett:2002nw}. In this paper, we take the same $SU(2)$ structure from \cite{Alday:2015jsa} and look for further generalizations of the solutions. In particular, we use the approach from \cite{Gauntlett:2004zh} to classify the families of solutions arising from a theory with a zero two-form potential $B$. We find that our six-dimensional space can be seen as a five-dimensional one which is orthogonal to the Killing vector. This five-dimensional space, in turn, can be seen as a product of a one-dimensional space and a four-dimensional space with a Kähler metric.

The plan for the paper is as follows. In section \ref{ERT}, we summarize Euclidean
Romans supergravity theory with 
a non-trivial Abelian R-symmetry gauge field. In section \ref{SecSU2}, we summarize the differential constraints imposed on the canonical $SU(2)$ structure of this Romans theory to ensure supersymmetric solutions. In section \ref{zerobfield} we take the B-field to be equal zero in the differential contraints forementioned, simplifying the problem. These new equations can now be solved in closed form. We end by classifying the families of solutions.

\section{Euclidean Romans supergravity}\label{ERT}

The bosonic fields of the six-dimensional Romans supergravity theory \cite{Romans:1985tw} consist of the metric, a scalar field $X=\exp(-\tfrac{\phi}{2\sqrt{2}})$ where $\phi$ is the dilaton, a two-form potential $B$, together with an $SO(3)_R\sim SU(2)_R$ R-symmetry gauge field $A^i$ with field strength $F^i=\diff A^i-\frac{1}{2}\varepsilon_{ijk}A^j\wedge A^k$, where $i=1,2,3$. 
Here we are working in a gauge in which the Stueckelberg one-form is zero, and we set 
the  gauge coupling constant to 1. 
  The Euclidean signature equations of motion are \cite{Alday:2014bta}
  \bea\label{FullEOM}
  \diff\left(X^{-1}*\dd X\right) &=& -  \left(\tfrac{1}{6}X^{-6}-\tfrac{2}{3}X^{-2}+\tfrac{1}{2}X^2\right)*1 \nn \\
&&-\tfrac{1}{8}X^{-2}\left(\tfrac{4}{9}B\wedge *B+F^i\wedge *F^i\right) + \tfrac{1}{4}X^4H\wedge *H ~,\nn\\
\diff\left(X^4 * H\right) &=& \tfrac{2\, \ii}{9}B\wedge B + \tfrac{\ii}{2}F^i\wedge F^i +\tfrac{4}{9}  X^{-2}*B~,\nn\\
D(X^{-2}*F^i) & = & - \ii F^i\wedge H~.
\eea
Here $H=\diff B$ and
$D\omega^i=\dd\omega^i - \varepsilon_{ijk}A^j\wedge \omega^k$ is the 
$SO(3)$ covariant derivative. Notice that the theory 
contains Chern-Simons-type couplings, that become purely imaginary in Euclidean signature.
The Einstein equation is
\bea\label{Einstein}
R_{\mu\nu} &=& 4X^{-2}\partial_\mu X\partial_\nu X + \left(\tfrac{1}{18}X^{-6}-\tfrac{2}{3}X^{-2}-\tfrac{1}{2}X^2\right) g_{\mu\nu} \nn \\ 
&& + \tfrac{1}{4}X^4\left(H^2_{\mu\nu}-\tfrac{1}{6}H^2g_{\mu\nu}\right)  +  \tfrac{2}{9}X^{-2}\left(B^2_{\mu\nu}-\tfrac{1}{8}B^2g_{\mu\nu}\right) \nn\\
&& +  \tfrac{1}{2}X^{-2}\left((F^i)^2_{\mu\nu}-\tfrac{1}{8}(F^i)^2g_{\mu\nu}\right)~,
\eea
where $B^2_{\mu\nu} = B_{\mu\rho} B_\nu{}^\rho$, $H^2_{\mu\nu}=H_{\mu\rho\sigma}H_{\nu}^{\ \rho\sigma}$.
  
A  solution is 
 supersymmetric provided there 
exists a non-trivial $SU(2)_R$ doublet of Dirac spinors $\epsilon_I$, $I=1,2$, satisfying 
the following Killing spinor  and dilatino equations
\bea
D_\mu \epsilon_I & =&  \tfrac{\ii}{4\sqrt{2}}  ( X + \tfrac{1}{3} X^{-3} ) \Gamma_\mu \Gamma_7 \epsilon_I - \tfrac{\ii}{24\sqrt{2}} X^{-1} B_{\nu\rho} ( \Gamma_\mu{}^{\nu\rho} - 6 \delta_\mu{}^\nu \Gamma^\rho ) \epsilon_I \nn  \\
&& - \tfrac{1}{48} X^2 H_{\nu\rho\sigma} \Gamma^{\nu\rho\sigma} \Gamma_\mu \Gamma_7 \epsilon_I
+ \tfrac{1}{16\sqrt{2}}X^{-1} F_{\nu\rho}^i ( \Gamma_\mu{}^{\nu\rho} - 6 \delta_\mu{}^\nu \Gamma^\rho ) \Gamma_7 ( \sigma_i )_I{}^J \epsilon_J ~,\nn \\\label{KSE}\\
0 & = & - \ii X^{-1} \partial_\mu X \Gamma^\mu \epsilon_ I + \tfrac{1}{2\sqrt{2}}  \left( X - X^{-3} \right) \Gamma_7 \epsilon_I + \tfrac{\ii}{24} X^2 H_{\mu\nu\rho} \Gamma^{\mu\nu\rho} \Gamma_7 \epsilon_I \nonumber \\
&&- \tfrac{1}{12\sqrt{2}} X^{-1} B_{\mu\nu} \Gamma^{\mu\nu} \epsilon_I - \tfrac{\ii}{8\sqrt{2}} X^{-1} F^i_{\mu\nu} \Gamma^{\mu\nu} \Gamma_7 ( \sigma_i )_I{}^J \epsilon_J~.\label{dilatino}
\eea
Here  $\Gamma_\mu$, $\mu=1,\ldots,6$, are taken to be Hermitian and generate the Clifford 
algebra $\mathrm{Cliff}(6,0)$ in an orthonormal frame. We have defined the chirality operator
 $\Gamma_7 = \ii \Gamma_{123456}$, which satisfies $(\Gamma_7)^2=1$. 
The covariant derivative acting on the spinor is $D_\mu\epsilon_I={\nabla}_\mu\epsilon_I+\frac{\ii}{2} A^i_\mu(\sigma_i)_I{}^J\epsilon_J$, where ${\nabla}_\mu=\partial_\mu+\frac{1}{4}\Omega_{\mu}^{\ \, \nu\rho}\Gamma_{\nu\rho}$ denotes the Levi-Civita 
spin connection while $\sigma_i$, $i=1,2,3$, are the Pauli matrices.

For simplicity we shall consider Abelian solutions in which 
$A^1_\mu=A^2_\mu=0$, and $A^3_\mu\equiv \mathcal{A}_\mu$, with field strength $\mathcal{F}\equiv \diff \mathcal{A}$. 
Also, as in \cite{Alday:2014bta}, we consider a ``real'' class of solutions
 for which $\epsilon_I$ satisfies the symplectic Majorana condition 
 $\varepsilon_{I}^{\ J}\epsilon_J = \mathcal{C}\epsilon_I^*\equiv \epsilon_I^c$, where 
 $\mathcal{C}$ denotes the charge conjugation matrix, satisfying $\Gamma_\mu^\mt=\mathcal{C}^{-1}\Gamma_\mu \mathcal{C}$. The bosonic fields 
 are all taken to be real, with the exception of the $B$-field which is purely imaginary and that will later be taken to be zero. 
 With these reality properties one can show that the Killing spinor equation (\ref{KSE}) and dilatino equation 
 (\ref{dilatino}) for $\epsilon_2$  are simply the charge conjugates of the corresponding 
 equations for $\epsilon_1$.
  In this way 
 we  effectively reduce to a single Killing spinor $\epsilon\equiv \epsilon_1$, with $SU(2)_R$ doublet $(\epsilon_1,\epsilon_2)=(\epsilon,\epsilon^c)$. 

\section{$SU(2)$ structure}\label{SecSU2}

Consider a Dirac spinor $\epsilon$ in six dimensions, such that $(\epsilon_1,\epsilon_2)=(\epsilon,\epsilon^c)$ 
solves  (\ref{KSE}) and (\ref{dilatino}) above. 
We may construct the following scalar bilinears
\bea
S &\equiv & \epsilon^\dagger\epsilon~, \qquad \tilde{S} \ \equiv \ \epsilon^\dagger\Gamma_7\epsilon~, \qquad 
f \ \equiv \ \epsilon^\mt \epsilon~.
\eea
We have chosen a basis for the gamma matrices in which they are purely imaginary and anti-symmetric, 
with charge conjugation matrix $\mathcal{C}=-\ii \Gamma_7$. 
A short computation reveals that
\bea
\diff (Xf) &=& -\ii ( Xf) \mathcal{A}~.
\eea
The integrability condition for this equation immediately implies 
$\mathcal{F}=\diff \mathcal{A}=0$ unless $f\equiv 0$ (notice that 
$X$ is nowhere zero). We will henceforth restrict our analysis to the 
case $f\equiv 0$, which is necessary for a non-trivial R-symmetry gauge field.\footnote{There are nevertheless interesting solutions 
for which $f\neq 0$. Indeed, we have constructed a 1/2 BPS solution in \cite{Alday:2014bta} lying in this class.} 

We may then write 
\bea\label{epm}
\epsilon &=& \epsilon_+ + \epsilon_-~,
\eea
where $-\Gamma_7\epsilon_\pm = \pm \epsilon_\pm$, and furthermore the condition $f\equiv 0$ allows us  to introduce
\bea\label{SU2rotation}
\epsilon_+ &=& \sqrt{S}\cos\alphaangle\, \eta_1~, \qquad \epsilon_- \ = \ \sqrt{S}\sin\alphaangle\, \eta_2^*~,
\eea
as done in \cite{Gauntlett:2004zh}. Here $\eta_1$, $\eta_2$ are two orthogonal unit norm chiral spinors, so that 
$\eta_1^\dagger\eta_1=\eta_2^\dagger\eta_2=1$ and $\eta_2^\dagger\eta_1=0$. 
These each define a canonical $SU(3)$ structure, and together determine a canonical  $SU(2)$ structure. 
Concretely, in six dimensions such a structure is specified by two one-forms
$K_1$, $K_2$ and a triplet of two-forms $J_i$, $i=1,2,3$, given by
\bea
K_1 - \ii K_2 &\equiv & -\frac{1}{2}\varepsilon^{\alpha\beta}\eta_\alpha^\mt\Gamma_{(1)}\eta_\beta~,\nn\\
J_i &\equiv  & -\frac{\ii}{2}\sigma_i^{\alpha\beta}\eta_\alpha^\dagger \Gamma_{(2)} \eta_\beta~.
\eea
Here we have introduced the notation $\Gamma_{(n)}\equiv \frac{1}{n!}\Gamma_{\mu_1\cdots \mu_n}\diff x^{\mu_1}\wedge \cdots \wedge \diff x^{\mu_n}$, 
where $x^\mu$  are local coordinates. We also define
\bea
\Omega & \equiv & J_2 + \ii J_1~, \qquad J \ \equiv \ J_3~.
\eea
The canonical $SU(2)$ structure is thus determined by $(K_1,K_2,J,\, \Omega)$. 
We note that $K_1$ and $K_2$ are orthonormal one-forms, and both are orthogonal 
to $J$ and $\Omega$, with $J\wedge \, \Omega=0$ and $2J\wedge J = \Omega\wedge\bar{\Omega}$. 

The $SU(2)$ structure $(S,\alphaangle,K_1,K_2,J,\, \Omega)$ that arises naturally from a supersymmetric 
solution is thus related to the canonical $SU(2)$ structure by the square norm $S$ and angle $\alphaangle$, 
via (\ref{SU2rotation}).

It was shown in \cite{Alday:2015jsa} that the differential contraints on this $SU(2)$ structure are given by
\bea
X^2S^2\sin^2 2\alphaangle \, \diff \sigma &=& -\tfrac{2\sqrt{2}}{3}X^{-1}S\cos 2\alphaangle \,  J 
- \ii X^4 S\sin 2\alphaangle\, K_1\hook *H_\perp \nn\\
&&  + \sqrt{2} XS (\cos 2\alphaangle \, \mathcal{F}_\perp + \tfrac{2}{3}\ii B_\perp)~,\nn\\
\diff (X^{-1}S\cos 2\alphaangle\, J) &=& -\tfrac{3}{2\sqrt{2}} \diff[(XS)^2 \diff\sigma] 
+ \ii XS \, \diff B_\perp \nn\\&&
+ \tfrac{\sqrt{2}}{3}\ii X^{-2}S\sin 2\alphaangle \left[K_1\hook *B_\perp 
- K_2\wedge B_\perp\right]~,\nn\\
\diff (X^{-1}SJ) &=& -\sqrt{2}S\sin 2\alphaangle\, J\wedge K_2 - \tfrac{3}{2\sqrt{2}}\cos 2\alphaangle \, \diff [(XS)^2 \diff \sigma]  \nn \\ && + \ii XS \cos 2\alphaangle\, \diff B_\perp
- \tfrac{1}{\sqrt{2}}X^{-2} S \sin 2\alphaangle\, \left[K_1\hook *\mathcal{F}_\perp - K_2\wedge\mathcal{F}_\perp\right]\, ,\nn\\
\diff (S\sin 2\alphaangle\, J\wedge K_2) &=& 0~,\nn\\
D_\perp(X^{-1}S\sin 2\alphaangle\, \Omega) &=& -\sqrt{2}S\Omega\wedge K_2~,\nn\\
S^2 J\wedge \diff\sigma &=& -\sqrt{2}S\cos 2\alphaangle (X+\tfrac{2}{3}X^{-3})\tfrac{1}{2}J\wedge J + 2S K_1\hook *\diff \alphaangle \nn\\
&& +\tfrac{1}{\sqrt{2}}X^{-1}SJ\wedge (\cos 2\alphaangle\, \diff\mathcal{A}_\perp + \tfrac{2}{3}\ii B_\perp)~,\nn\\
S^2 \Omega\wedge \diff\sigma & =&   -2\ii S \diff \alphaangle\wedge K_2 
\wedge \Omega  + \tfrac{1}{\sqrt{2}}X^{-1}S\Omega\wedge (\cos 2\alphaangle\, 
\diff\mathcal{A}_\perp + \tfrac{2}{3}\ii B_\perp)~,\nn\\
0 &=& X^4 K_2 \hook \diff (X^{-3}S\sin 2\alphaangle) + \sqrt{2}S(X^2-\tfrac{2}{3}X^{-2})  \nn\\
&&+\tfrac{1}{\sqrt{2}}S J\hook
(\mathcal{F}_\perp + \tfrac{2}{3}\ii \cos 2\alphaangle\, B_\perp)~,\label{diffconstraints}
\eea
where fields were divided into components parallel and perpendicular to $K_1$, so that $K_1\hook\mathcal{F}_\perp=0$. It is also shown in \cite{Alday:2015jsa} that, when supplemented by 
\bea
\label{extra}
&&0 \ =\ X^4 S\sin 2\alphaangle \, \diff \sigma \wedge (K_1\hook *\, \ii  H_\perp) + \diff \left[\frac{X^4}{S\sin 2\alphaangle}K_1\hook * \diff (X^{-2} S \sin 2\alphaangle \, K_2)\right]
 \nn\\
 && +\tfrac{2}{9}B_\perp\wedge B_\perp + \tfrac{1}{2}\mathcal{F}_\perp\wedge \mathcal{F}_\perp - 
 \tfrac{4 }{9}X^{-2}K_1\hook  *\left[\frac{3}{\sqrt{2}S \sin 2\alphaangle}\diff (XS) + X^{-2}K_2\right]~, \nn \\
\eea
solving these equations is equivalent to finding supergravity solutions to Euclidean Romans theory.

\section{Classifying solutions with a zero B-field} \label{zerobfield}

Briefly, the idea is to take Euclidean Romans supergravity theory and set the two-form potential $B$ to be zero. This leads us to a simpler set 
of equations of motion, dilatino and Killing spinor equations.

We analyzed possible solutions to Romans supergravity in previous papers \cite{Alday:2014bta,Alday:2015jsa}, and a particular case where $B=0$ was discussed in \cite{Alday:2014fsa}. One can expect, however, to find more possible solutions to this case. By following the procedure adopted in \cite{Gauntlett:2004zh}, here we classify these families of solutions.

\subsection{Taking $B=0$}

Taking the two-form potential $B=0$ in \eqref{diffconstraints} and \eqref{extra} implies
\bea
X^2S^2\sin^2 2\alphaangle \, \diff \sigma &=& -\tfrac{2\sqrt{2}}{3}X^{-1}S\cos 2\alphaangle \,  J 
 + \sqrt{2} XS \cos 2\alphaangle \, \mathcal{F}_\perp~,\label{eq1}\\
\diff (X^{-1}S\cos 2\alphaangle\, J) &=& -\tfrac{3}{2\sqrt{2}} \diff[(XS)^2 \diff\sigma]~,\label{eq2}\\
\diff (X^{-1}SJ) &=& -\sqrt{2}S\sin 2\alphaangle\, J\wedge K_2 - \tfrac{3}{2\sqrt{2}}\cos 2\alphaangle \, \diff [(XS)^2 \diff \sigma]  \nn \\ 
&& - \tfrac{1}{\sqrt{2}}X^{-2} S \sin 2\alphaangle\, \left[K_1\hook *\mathcal{F}_\perp - K_2\wedge\mathcal{F}_\perp\right]\, ,\label{eq3}\\
\diff (S\sin 2\alphaangle\, J\wedge K_2) &=& 0~,\label{eq4}\\
D_\perp(X^{-1}S\sin 2\alphaangle\, \Omega) &=& -\sqrt{2}S\Omega\wedge K_2~,\label{eq5}\\
S^2 J\wedge \diff\sigma &=& -\sqrt{2}S\cos 2\alphaangle (X+\tfrac{2}{3}X^{-3})\tfrac{1}{2}J\wedge J + 2S K_1\hook *\diff \alphaangle \nn\\
&& +\tfrac{1}{\sqrt{2}}X^{-1}SJ\wedge \cos 2\alphaangle\, \diff\mathcal{A}_\perp ~,\label{eq6}\\
S^2 \Omega\wedge \diff\sigma & =& \tfrac{1}{\sqrt{2}}X^{-1}S\Omega\wedge \cos 2\alphaangle\, 
\diff\mathcal{A}_\perp  -2\ii S \diff \alphaangle\wedge K_2 
\wedge \Omega  ~,\label{eq7}\\
0 &=& X^4 K_2 \hook \diff (X^{-3}S\sin 2\alphaangle) + \sqrt{2}S(X^2-\tfrac{2}{3}X^{-2})  \nn\\
&&+\tfrac{1}{\sqrt{2}}S J\hook
\mathcal{F}_\perp~,\label{eq8}
\eea
and
\begin{equation}
\label{eq9}
0 \ = \diff \left[\frac{X^4}{S\sin 2\alphaangle}K_1\hook * \diff (X^{-2} S \sin 2\alphaangle \, K_2)\right]+ \tfrac{1}{2}\mathcal{F}_\perp\wedge \mathcal{F}_\perp~.
\end{equation}
Solving these equations is sufficient to ensure we have a supersymmetric solution to the Euclidean equations of motion of Romans theory in this particular case. 

It is also worth recalling from \cite{Alday:2015jsa} that the one-form $K_1$ can be written as 
\begin{equation}
K_1=S \sin{2\alphaangle}(\dd \psi +\sigma)~, \nn
\end{equation}
where $\partial_\psi$ is the supersymmetric Killing vector that preserves all the structure, and also that part of $\mathcal{F}$ perpendicular to $K_1$ can be written as
\begin{equation}
\mathcal{F}_\perp = -\sqrt{2}X S \cos{2\alphaangle}\dd\sigma + \dd\mathcal{A}_\perp~. \label{fperp}
\end{equation}

In order to analyse these equations, we will need to break the derivatives (and the remaining of each equation) further into components. We start by defining a radial coordinate in the $K_2$ direction, 
this will given by 
\begin{equation}
 \rho = X S~,~~~ \text{such that} ~~~\dd \rho = \dd(XS)~.
\end{equation}

The exterior derivative can be written as
\begin{equation}
 \dd = \dd\psi\wedge \partial_\psi ~+~\dd\rho\wedge \partial_\rho~+~\dd_4~,
\end{equation}
i.e., in the $K_1$, $K_2$ and $M_4$ directions.

Another term that requires attention is the one-form $\sigma$. For instance, one may consider
\begin{equation}
 \sigma=\sigma_4~+~\sigma_\rho \dd\rho~,
\end{equation}
where $\sigma_4$ is the $\sigma$ component in the direction that is both perpendicular to $K_1$ and $K_2$. Its derivative is then given by
\begin{equation}
 \dd \sigma = \dd_4\sigma_4~+~\dd\rho\wedge(\partial_\rho\sigma_4)~+~(\dd_4\sigma_\rho)\wedge\dd\rho.
\end{equation}
Notice however that, if we reparametrise $\psi$ (that enters in the definition of the one-form $K_1$), one can make a gauge choice of shifting 
it in such a way that 
\begin{equation}
 \psi \longrightarrow \psi - \int_\rho \sigma_\rho(y,x^i) \dd y~,
\end{equation}
which leads to 
\begin{equation}
 \dd \psi \longrightarrow \dd \psi -\sigma_\rho \dd\rho - \dd_4\int_\rho \sigma_\rho (y, x^i) \dd y~.
\end{equation}
This way, in writing down $\dd\psi+\sigma$, we have $\sigma_\rho\dd\rho$ being cancelled, and we can simply write
\begin{equation}
 \dd\psi+\sigma \longrightarrow \dd\psi+ \sigma_4 - \dd_4\int_\rho \sigma_\rho (y, x^i) \dd y~,
\end{equation}
and relabel
\begin{equation}
  \sigma_4 - \dd_4\int_\rho \sigma_\rho (y, x^i) \dd y \longrightarrow \sigma_4~,
\end{equation}
as both terms are in the $\dd_4$ direction. This way, we 
are free to make a choice where $\sigma\equiv\sigma_4$. Notice however that we still 
have to consider $\dd \sigma = \dd_4\sigma_4+\dd\rho\wedge(\partial_\rho\sigma_4)$.


\subsection{Conditions for a supersymmetric $M_6$}


Notice that once we take the $B$-field to be zero, we immediately get $B_1=0$, this gives us
\begin{equation}
 \frac{3}{\sqrt{2}S\sin 2\alphaangle}\diff (XS) + X^{-2} K_2=0~.
\end{equation}
As we defined, $\dd(XS)=\dd\rho$ (now confirming that $K_2$ is in fact in the $\dd\rho$ direction), indeed,
\begin{equation}
 K_2 = - \frac{3 X^2}{\sqrt{2}S\sin 2\alphaangle}\dd \rho~. 
\end{equation}
In \eqref{eq9}, notice that the first term is zero, and it simply reduces to 
\begin{equation}
\mathcal{F}_\perp \wedge\mathcal{F}_\perp=0~.
\end{equation}
Equation \eqref{eq1} may be used to eliminate the flux $\mathcal{F}_\perp$ in terms of the $SU(2)$ structure (one should consider $\cos{2\alphaangle}\equiv 0$ as a separate case) such that 
\begin{equation}
\mathcal{F}_\perp=\frac{2}{3X^2}J+\frac{\rho \sin^2{2\alphaangle}}{\sqrt{2}\cos{2\alphaangle}}\dd\sigma~. \label{2.2}
\end{equation}

Let us reparametrize $J$ and $\Omega$ by introducing
\begin{equation}
 \hat{J}=X^2~J~, ~~~ \text{and} ~~~ \hat{\Omega} = X^2~\Omega~.
\end{equation}
Then from equation \eqref{eq4}, one reads
\begin{equation}
\dd(\hat{J}\wedge \dd \rho)=0~.
\end{equation} 
which implies
\begin{align}
 \dd_4 \hat{J} &=0~, \label{dJ} \\
 \partial_\psi \hat{J}&=0~, \label{dpJ}
\end{align}
 
Next, equation \eqref{eq2} reads 
\begin{equation}
\dd\left(\frac{\rho \cos{2\alphaangle}}{X^4}\hat{J}\right)=-\frac{3}{\sqrt{2}}\rho~\dd\rho\wedge\dd\sigma~.
\end{equation}
This is equivalent to
\begin{equation}
 \dd_4\left(\frac{\cos{2\alphaangle}}{X^4}\right)=0~, \label{frho}
\end{equation}
so that we can say $\cos{2\alphaangle}=X^4\lambda(\rho)$. We then get an equivalent to equation (2.36) in \cite{Gauntlett:2004zh}, namely
\begin{equation}
\partial_\rho(\rho\lambda(\rho)\hat{J})=-\frac{3}{\sqrt{2}}\rho\dd_4\sigma~. \label{2.8}
\end{equation}
Notice that from this equation also follows that $\dd_4 \sigma$ has no components proportional to $\Omega$ (but it still could have an anti-self-dual part).

Similarly, equation \eqref{eq5} reads
\begin{equation}
 \dd_4 \left(\frac{\sin{2\alphaangle}}{X^4}\hat{\Omega}\right) = - \ii \mathcal{A}_\perp
  \wedge \frac{\sin{2\alphaangle}}{X^4} \hat{\Omega}~, \label{2.10}
\end{equation} 
and
\begin{equation}
  \partial_\rho \left(\frac{\rho\sin{2\alphaangle}}{X^4}\hat{\Omega}\right)  =\frac{3}{\sin{2\alphaangle}}\hat{\Omega}~. \label{2.11}
\end{equation} 
Here again we have used gauge freedom to remove the part of $\mathcal{A}_\perp$ proportional to $\dd\rho$, so that $\mathcal{A}_\perp=\mathcal{A}_4$. These equations imply that the geometry at constant $\rho$ (and $\psi$) is (conformally) K\"ahler, with K\"ahler metric $\hat{g}_4$ associated to $\hat{J}$ and $\hat{\Omega}$. Moreover, since the derivative of $\hat{\Omega}$ in the $\rho$ direction is proportional to $\hat{\Omega}$, this shows that the associated complex structure $\hat{I}$ is independent of $\rho$, $\partial_\rho\hat{I}=0$. Since
%
\begin{equation}
 \dd_4 \hat{\Omega} = \ii \hat{P}\wedge\hat{\Omega},
\end{equation}
where $\hat{P}$ is the canonical Ricci one-form potential, we identify
\begin{equation}
\mathcal{A}_\perp=-\hat{P}-\hat{I}\cdot\dd_4 \log{\left(\frac{\sin{2\alphaangle}}{X^4}\right)}~.
\end{equation}
Note that this can be rewritten as 
\begin{equation}
\mathcal{A}_\perp=-\hat{P}-\hat{I}\cdot\dd_4 \log{\tan{2\alphaangle}}~. \label{2.14}
\end{equation}

Next, we turn to equation \eqref{eq6}. Multiplying it by $X^2$ and substituting from  \eqref{fperp}, this reads
\begin{equation}
X^2S^2\sin^2{2\alphaangle}\Omega\wedge\dd\sigma = -2\ii S \dd \alphaangle \wedge K_2\wedge \Omega+ \frac{1}{\sqrt{2}}XS\Omega\wedge \cos{2\alphaangle}\mathcal{F}_\perp~.
\end{equation}
Subtracting $\tfrac{1}{2}\Omega$ times \eqref{eq1} from this previous equation gives
\begin{equation}
\frac{1}{2}\rho^2 \sin^2{2\alphaangle} \omega\wedge\dd\sigma = -2\ii S\dd\alphaangle \wedge K_2\wedge\Omega~.
\end{equation}
And therefore
\begin{equation}
\Omega\wedge\dd_4\sigma=0~,
\end{equation}
and
\begin{equation}
\Omega\wedge\left(\partial_\rho\sigma +\frac{6\sqrt{2}X^4}{\rho^2\sin^3{2\alphaangle}}\ii~\dd_4\alphaangle\right)=0~.
\end{equation}

This implies that the one-form in brackets is a $(1,0)-$form, and hence
\begin{equation}
\partial_\rho\sigma  = - \frac{6\sqrt{2}X^4}{\rho^2\sin^3{2\alphaangle}}\hat{I}\dot \dd_4\alphaangle~. \label{2.20}
\end{equation}

Next we turn to equation \eqref{eq6}. One finds that the component of this equation in the $\dd \rho$ direction is precisely equivalent to equation \eqref{2.20}. The remainder of equation \eqref{eq6} is equivalent to
\begin{equation}
\frac{1}{2}S^2\sin^2{2\alphaangle}J\hook\dd_4\sigma = -\sqrt{2}\rho\cos{2\alphaangle}+\frac{2\sqrt{2}\rho^2\sin{2\alphaangle}}{3X^4}\partial_\rho\alphaangle~. \label{2.21}
\end{equation}

Finally we turn to the scalar equation \eqref{eq8}. After a computation, one remarkably finds precisely equation \eqref{2.21} plus a (generically) non-zero function times $\partial_\rho(\rho\lambda(\rho))$. One concludes that
\begin{equation}
\partial_\rho(\rho\lambda(\rho))=0 \longrightarrow \lambda(\rho)=\frac{c}{\rho}~, \label{2.22}
\end{equation}
where $c$ is an integration constant. One can check that equation \eqref{2.22} is also true for the four-parameter family of BPS black hole solutions discussed in \cite{Alday:2014fsa}, a highly non-trivial check. We have thus solved these equations for one of the functions in the problem.  
Notice that now one can write
\begin{equation}
\rho = cX^4 \sec{2\alphaangle}~.
\end{equation}
We thus really have only one free function in the problem, and we can take it to be $X$.

We have now analysed all the content of all the equations, apart from equations \eqref{eq3} and \eqref{eq9}. After quite a lengthy calculation, and using many of the equations above, one can show that the $\dd\rho$ component of equation \eqref{eq3} is precisely equivalent to \eqref{2.21}. Notice that the anti-self-dual part of $\mathcal{F}_4$ enters, which is related to $(\dd_4\sigma)^{-}$ via  \eqref{2.2}, but this combines with $\dd_4\sigma$, and in the end only the self-dual part of the it remains, and it is proportional to $J$, as $\Omega\wedge\dd_4\sigma=0$. The remainder of equation \eqref{eq3} is easier to compute, and one finds an equivalent to \eqref{2.20}. Thus \eqref{eq3} is implied by all the other equations, and hence imposes nothing new.

It thus remains only to impose the equation of motion \eqref{eq9}. The two components read
\begin{equation}
\partial_\rho\sigma\wedge\mathcal{F}_4=0~. \label{2.24}
\end{equation}
where from \eqref{2.2}, we have 
\begin{equation}
\mathcal{F}_4=\frac{2}{3X^2}J+\frac{\rho \sin^2{2\alphaangle}}{\sqrt{2}\cos{2\alphaangle}}\dd_4\sigma~, 
\end{equation}
together with the scalar equation 
\begin{equation}
 \|(\dd_4\sigma)^{-}\|= \frac{2\cos{2\alphaangle}}{\rho\sin^2{2\alphaangle}}
 \left[ \frac{2\rho\sin{2\alphaangle}}{3X^2\cos{2\alphaangle}}\partial_\rho\alphaangle
 + \frac{1}{X^2}\left(\frac{2}{3}-X^4\right)\right]~.
\end{equation}

We conclude by noting that a few equations are redundant. First \eqref{2.20} is precisely the $\dd\rho$ component of $\dd$\eqref{2.14}. Here we have the second equation
\begin{equation}
\dd\mathcal{A}_\perp = \frac{2}{3X^4}\hat{J}+\frac{\rho}{\sqrt{2}}(\cos{2\alphaangle}+\sec{2\alphaangle})\dd\sigma~,
\end{equation}
which may be combined with \eqref{2.24} to obtain an Einstein-like equation (involving the Ricci form of the K\"ahler metric). To see this, recall that $\hat{\mathcal{P}}=\tfrac{1}{2}\hat{I}\dot\dd_2\log{\sqrt{\det{\hat{g}}}}$. But since also $\hat{\Omega}\wedge\hat{\Omega}=4\hat{\text{vol}}=2\hat{J}\wedge\hat{J}$ is automatically true, it follows by taking $\partial_\rho$ that
\begin{equation}
\hat{J}\hook\partial_\rho\hat{J}=\hat{\Omega}\hook\partial_\rho\hat{\Omega}
\end{equation}
is an identity. Using this, one can check that equations \eqref{2.8} and \eqref{2.11} in fact imply equation \eqref{2.21}. The latter is hence implied by the other equations, and it is also  redundant.


\subsubsection{Summary}
We can now put together the necessary and sufficient conditions to have a supersymmetric solution. The metric is given by
\begin{equation}
 \dd s_6^2= K_1^2+K_2^2+g_{SU(2)}~,
\end{equation}
where now $g_{SU(2)}$ is a conformally K\"ahler manifold. We can rewrite it as
\begin{equation}
  \dd s_6^2= X^{-2}\left(\rho^2\sin^2{2\alphaangle}(\dd\psi+\sigma)^2+\frac{9X^8}{2\rho^2\sin^2{2\alphaangle}}\dd\rho^2
 + \hat{g}_{ij}\dd x^i \dd x^j\right)~, \label{kahler}
\end{equation}
where $\hat{g}$ is a one-parameter family of K\"ahler metrics depending only on $\rho$, for which the complex strucuture $\hat{I}$ is independent of $\rho$. The vector $\partial_\psi$ is Killing, and preserves all of the structure. The functions $X$ and $\alphaangle$ are related by
\begin{equation}
X^4=\frac{\rho}{c}\cos{2\alphaangle}~, \label{2.31}
\end{equation}
where c is a non-zero constant, so that we may substitute
\begin{equation}
\sin^2{2\alphaangle}=1-\frac{c^2X^8}{\rho^2}~.
\end{equation}
The evolution equations for the K\"ahler structure are 
\begin{equation}
\partial_\rho\hat{J}=-\frac{3}{\sqrt{2}c}\rho\dd_4\sigma~ \label{2.33}
\end{equation}
and
\begin{equation}
  \partial_\rho \left(\tan{2\alphaangle}\hat{\Omega}\right)  =\frac{3}{c\sin{2\alphaangle}}\hat{\Omega}~. \label{2.34}
\end{equation} 
Notice that $\Omega\wedge\dd_4\sigma=0$ is consistent with equation \eqref{2.33} and $\hat{J}$ must remain type $(1,1)$ as the complex structure is independent of $\rho$.

From \eqref{2.14}, we have the Einstein-like equation
\begin{equation}
\mathcal{\hat{R}}\equiv\dd_4\mathcal{\hat{P}}=-\frac{2}{3X^4}\hat{J}-\frac{\rho}{\sqrt{2}}(\cos{2\alphaangle}+\sec{2\alphaangle})\dd_4\sigma -\dd_4\cdot\hat{I}\cdot\dd_4\log{\tan{2\alphaangle}}~, \label{2.35}
\end{equation}
together with
\begin{equation}
\partial_\rho\sigma=-\frac{6\sqrt{2}X^4}{\rho^2\sin^3{2\alphaangle}}\hat{I}\cdot \dd_4\alphaangle~.
\end{equation}
Finally, we must impose the B-field equation of motion components
\begin{equation}
\partial_\rho\sigma\wedge\left[\frac{2}{3X^4}\hat{J}+\frac{\rho\sin^2{2\alphaangle}}{\sqrt{2}\cos{2\alphaangle}}\dd_4\sigma\right]=0~,
\end{equation}
and 
\begin{equation}
 \|(\dd_4\sigma)^{-}\|= \frac{2\cos{2\alphaangle}}{\rho\sin^2{2\alphaangle}}
 \left[ \frac{2\rho\sin{2\alphaangle}}{3X^2\cos{2\alphaangle}}\partial_\rho\alphaangle
 + \frac{1}{X^2}\left(\frac{2}{3}-X^4\right)\right]~. \label{2.38}
\end{equation}
The norm here is with respect to $g_4$ (rather than $\hat{g}_4$). Notice that remarkably the supersymmetry equations above are almost exactly the same (essentially up to numerical factors) to the equations in \cite{Gauntlett:2004zh}.

\subsection{Complex $M_6$ (Setting $\dd_4\alphaangle =0$)}

In \cite{Gauntlett:2004zh} the equations of this form were solved in closed form, with the additional assumption of $\dd_4\alphaangle =0$, leading to new solutions. It is then natural, due to the similarity of the system, to make the same assumption here.

In order to have a six-dimensional complex manifold with Hermitian metric, we require the three-form 
given by $\Omega_{(3)}=\Omega\wedge(K_1+\ii K_2)$ to have a derivative in the form
\begin{equation}
 \dd \Omega_{(3)}=A\wedge\Omega_{(3)}+v\wedge\Omega\wedge(K_1-\ii K_2),
\end{equation}
with $v=0$. This restriction will imply that $\dd\sigma\equiv\dd_4\sigma$, and implies that 
$\dd_4X=\dd_4\alphaangle=\dd_4S=0$. From this, one can deduce that 
\begin{equation}
 \hat{P}=\mathcal{A}_4.
\end{equation}

Next, we may look at \eqref{2.35}, which reads
\begin{equation}
\mathcal{\hat{R}}=-\frac{2}{3X^4}\hat{J}-\frac{\rho}{\sqrt{2}}(\cos{2\alphaangle}+\sec{2\alphaangle})\dd_4\sigma~. \label{3.2}
\end{equation}
The Ricci scalar of the K\"ahler metric $\hat{g}_4$ is $\hat{R}=\hat{J}^{ij}\hat{\mathcal{R}}_{ij}$, so that using \eqref{2.21}, we compute
\begin{equation}
\hat{R}=-\frac{4}{3X^4}-\rho(\cos{2\alphaangle}+\sec{2\alphaangle})\left(\frac{2\cos{2\alphaangle}}{\rho\sin^2{2\alphaangle}}-\frac{4}{3X^4\sin{2\alphaangle}}\partial_\rho\alphaangle\right)~.
\end{equation}
Since the right hand side is a function only of $\rho$, we deduce that $\dd_4\hat{R}=0$, and $\hat{g}_4$ is a constant scalar curvature K\"ahler metric (for fixed $\rho$).

We may similarly compute $\hat{R}_{ij} \hat{R}^{ij}= \hat{\mathcal{R}}_{ij} \hat{\mathcal{R}}^{ij}$ from equation \eqref{3.2}. Using again \eqref{2.21} to compute $\hat{J}\hook\dd_4\sigma$, and \eqref{2.38} to compute $\|(\dd_4\sigma)^{-}\|^2$, the right hand side is again a function only of $\rho$, and we deduce that 
\begin{equation}
 \dd_4(\hat{R}_{ij} \hat{R}^{ij})=0.
\end{equation}

It follows that at fixed $\rho$, the Ricci tensor $\hat{R}_{ij}$ has two pairs of constant eigenvalues. If these eigenvalues are the same, this is a K\"ahelr-Einstein metric, while if they are distinct and $M_4$ is compact, then the Goldberg conjecture implies that $M_4$ is locally a product of two Riemann surfaces of (distinct) constant curvature.

We shall consider both cases separately.

\subsubsection{K\"ahler-Einstein base solutions}

For a K\"ahler -Einstein metric $\mathcal{\hat{R}}\propto\hat{J}$, with the constant of proportionality depending only on $\rho$. Thus 
$\dd_4\sigma$ is also proportional to $\hat{J}$. One checks that there are no solutions with $\dd_4\sigma=0$, so, without loss of generality, we set
\begin{equation}
 \dd_4\sigma=\tilde{J}~, ~~~\hat{J}=F(\rho)\tilde{J}~,
\end{equation}
where $\partial_\rho\tilde{J}=0$. Thus the rescaled K\"ahler metric $\tilde{J}$ is independent of $\rho$, and the K\"ahler-Einstein condition reads
\begin{equation}
 \mathcal{\hat{R}}=\mathcal{\tilde{R}}=\kappa\tilde{J}~,
\end{equation}
where $\kappa\in\mathbb{R}$ is a constant. Solving  first equation \eqref{2.33}, we find
\begin{equation}
 F(\rho)=a-\frac{3\rho^2}{2\sqrt{2}c}~,
\end{equation}
where $a$ in an integration constant. Substitutuing this into \eqref{3.2}, and $X$ in \eqref{2.31}, one can find $\alphaangle$, given by
\begin{equation}
 \cos(2\alphaangle)=\left(-1+\sqrt{1-\frac{4\sqrt{2}ac}{3\kappa^2}}\right)\frac{\kappa}{\sqrt{2}\rho}~.
\end{equation}
Notice that at this point, all the functions have been completely determined. Next, solving \eqref{2.34}, we can write $a$ and $c$ in terms of $\kappa$ (where 
$\hat{\Omega}=F(\rho)\tilde{\Omega}$)
\begin{equation}
 a=-\frac{3\kappa}{4}~,~~~c=-\frac{\kappa}{\sqrt{2}}~.
\end{equation}
It follows that $X\equiv1$. Finally, notice that $(\dd_4\sigma)^{-}=0$, and one can check that the right hand side of the equation \eqref{2.38} is in fact zero. At this point we have solved all the equations. 

The final solution is therefore given by
\begin{equation}
 F(\rho)=\frac{3\rho^2}{2\kappa}-\frac{3\kappa}{4}~,~~~\cos(2\alphaangle)=-\frac{\kappa}{\sqrt{2}\rho}~,~~~X\equiv1~.
\end{equation}
The six-dimensional metric is 
\begin{equation}
 \dd s_6^2=\frac{9}{2(\rho^2-\tfrac{\kappa^2}{2})}\dd\rho^2+\left(\rho^2-\frac{\kappa^2}{2}\right)(\dd\psi+\sigma)^2
 +\frac{3}{2\kappa}\left(\rho^2-\frac{\kappa^2}{2}\right)\tilde{g}_4~, \label{mkebz}
\end{equation}
where $\dd\sigma=\tilde{J}$, and $\tilde{g}_4$ is a constant (in $\rho$) K\"ahler-Einstein metric with $\mathcal{\tilde{R}}=\kappa\tilde{J}$. 
The gauge field $\mathcal{A}$ has $\dd\mathcal{A}=-\dd\mathcal{\tilde{R}}$, so that $\mathcal{A}$ is a connection on the canonical bundle of $M_4$. 

The $\rho$ coordinate in the metric \eqref{mkebz} is somewhat peculiar. A better system of coordinates is set by making the change
\begin{equation}
 r^2\equiv\rho^2-\frac{\kappa^2}{2}~.
\end{equation}
The metric then becomes
\begin{equation}
 \dd s^2_6=\frac{9}{\kappa^2+2r^2}\dd r^2+r^2\left[(\dd \psi+\sigma)^2+\frac{3}{2\kappa}\hat{g}_4\right]~.
\end{equation}
This is simply the hyperbolic cone over a regular Sasaki-Einstein manifold. Notice that the full gauge field $\mathcal{F}=0$. Thus, this solution 
was known.

\subsubsection{Product of two Riemann surfaces base solutions}

Analogous to the K\"ahler-Einstein solutions, we can consider the metric where
\begin{equation}
 \hat{J}=F_1(\rho)\tilde{J}_1+F_2(\rho)\tilde{J}_2~,
\end{equation}
with $F_1(\rho)$ and $F_2(\rho)$ depending only on $\rho$. Then
we also have $\dd_4\sigma$ given by
\begin{equation}
 \dd_4\sigma=c_1\tilde{J}_1+c_2\tilde{J}_2~,
\end{equation}
with the factors $c_1$ and $c_2$ being constants. Here the two-forms 
$\tilde{J}_1$ and $\tilde{J}_2$ are such that $\partial_\rho\tilde{J}_1= \partial_\rho\tilde{J}_2=0$. Again the rescaled K\"ahler metric is independent of $\rho$ and we can write
\begin{equation}
 \hat{\mathcal{R}}=k_1\tilde{J}_1+k_2\tilde{J}_2~,
\end{equation}
where $k_i \in \mathbb{R}$ are constants.

Solving equation \eqref{2.33}, we find
\begin{equation}
 F_1(\rho)= -\frac{3\rho^2~c_1}{2\sqrt{2}c}-a ~~~\text{and}~~~ F_2(\rho)= -\frac{3\rho^2~c_2}{2\sqrt{2}c}-b~, 
\end{equation}
where $a$ and $b$ are integration constants. We can now substitute this into \eqref{3.2}, with $X$ given 
by \eqref{2.31}. Notice that there are two equations for $\alphaangle$. We find
\begin{equation}
 \cos{2\alphaangle}=\frac{4~a~c}{-3k_1~\rho+\sqrt{3\rho^2(3k_1^2-4\sqrt{2}a~c~c_1)}},
\end{equation}
and a constraint for $b$, given by
\begin{equation}
 b=6\left(\frac{2\sqrt{2}a^2~c~c_2~\rho^2}{\left(3k_1~\rho-\sqrt{3\rho^2(3k_1^2-4\sqrt{2}a~c~c_1)}\right)^2}+
 \frac{a~k_2~\rho}{3k_1~\rho -\sqrt{3\rho^2(3k_1^2-4\sqrt{2}a~c~c_1)}}\right)~.
\end{equation}

Now solving equation \eqref{2.34} order by order, one also finds constraints to $c$, $a$ and $c_2$, namely
\begin{equation}
 c=-\frac{k_1}{\sqrt{2}c_1}~, ~~~~~ a=-\frac{3k_1}{4}~, ~~~~~c_2=c_1\frac{k_2}{k_1}~.
\end{equation}
We then check equation \eqref{2.38} and confirm that it holds.

The final solution is a function of the parameters left, i.e., $k_1$, $k_2$ and $c_1$, and it is given by
\begin{equation}
 F_1(\rho)= \frac{3~c_1^2~\rho^2}{2~k_1}-\frac{3~k_1}{4} ~~~\text{and}~~~ F_2(\rho)= \frac{3~c_1~c_2~k_2~\rho^2}{2~k_1}-\frac{3~k_1~c_2}{4~c_1}~, 
\end{equation}
and our starting functions
\begin{equation}
 X=1~, ~~~ \text{and}~~~ \cos{2\alphaangle}=-\frac{k_1}{\sqrt{2}~c_1~\rho}~.
\end{equation}
The six-dimensional metric is
\begin{align}
 \dd s_6^2=~&\rho^2\left(1-\frac{k_1^2}{2~c_1~\rho}\right)(\dd\psi+\sigma)^2+\frac{9}{2~\rho^2}\left(1-\frac{k_1^2}{2~c_1~\rho}\right)^{-1} \dd\rho^2 \nn \\
 &+\left(\frac{3~c_1^2~\rho^2}{2~k_1}-\frac{3~k_1}{4}\right)\dd\tilde{s}^2(C_{k_1}) + 
 \left(\frac{3~c_1~c_2~k_2~\rho^2}{2~k_1}-\frac{3~k_1~c_2}{4~c_1}\right)\dd\tilde{s}^2(C_{k_2})~,
\end{align}
where $\dd \tilde{s}^2(C_{k_i})$ are the metrics on a torus ($C_0\equiv T^2$), or sphere ($C_1\equiv S^2$) 
or a hyperbolic space ($C_{-1}\equiv H^2$).

One can reparametrise the metric making $\dd s_6^2\longrightarrow k_1\dd s_6^2$, so that we get 
\begin{align}
 \dd s_6^2=~&\rho^2k_1\left(1-\frac{k_1^2}{2~c_1~\rho}\right)(\dd\psi+\sigma)^2+\frac{9k_1}{2~\rho^2}\left(1-\frac{k_1^2}{2~c_1~\rho}\right)^{-1} \dd\rho^2 \nn \\
 &+\left(\frac{3~c_1^2~\rho^2}{2}-\frac{3~k_1^2}{4}\right)\dd\tilde{s}^2(C_{k_1}) + 
 \left(\frac{3~c_1~c_2~\rho^2}{2}-\frac{3~k_1^2~c_2}{4~c_1}\right)\dd\tilde{s}^2(C_{k_2})~,
\end{align}
which can be rewritten as
\begin{align}
 \dd s_6^2=~&\rho^2k_1\left(1-\frac{k_1^2}{2~c_1~\rho}\right)(\dd\psi+\sigma)^2+\frac{9k_1}{2~\rho^2}\left(1-\frac{k_1^2}{2~c_1~\rho}\right)^{-1} \dd\rho^2 \nn \\
 &+\left(\frac{3~c_1^2~\rho^2}{2}-\frac{3~k_1^2}{4}\right)\dd\tilde{s}^2(C_{k_1}) + 
 \left(\frac{3~c_1^2\rho^2}{2}-\frac{3~k_1^2}{4}\right)\frac{k_2}{k_1}\dd\tilde{s}^2(C_{k_2})~.
\end{align}

Notice that the factor $\tfrac{k_1}{k_2}$ changes the curvature of $C_{k_2}$ to $C_{k_1}$, what we see therefore is a reduction back to the case where the base is simply a K\"ahler-Einstein 
manifold. We conclude that a K\"ahler-Einstein 
manifold is the most general solution to Romans supergravity with zero $B$ field (and $\dd_4\alphaangle=0$), completely classifying solutions of this type.

\section{Conclusion}

We have analysed the differential contraints in the $SU(2)$ structure obtained to ensure a supersymmetric solution to Euclidean Romans gravity theory when we set the two-form potential $B$ to be zero. We were able to solve them analytically and were able to see that the most natural global structure for our six-dimensional space is a $(\psi, \rho)$-holomorphic complex cone bundle over a Kähler base $M_4$, where $\psi$ is the Killing vector direction and $\rho$ an equivallent of a radial direcion. We then assumed the Goldberg conjecture to be true, and found that the $M_4$ manifold has to be either Kähler-Einstein, or a product of two Riemann surfaces.

We found that the only type of product of Riemann surfaces our $SU(2)$ structure allowed is that of two surfaces with the same curvature, returning therefore to the Kähler-Einstein case. This proves that the complex cone holomorphic bundle over a Kähler-Einstein base is the most general type of solution one can get for Romans supergravity theory with a zero B-field.

\section*{Acknowledgment}

The work of the author is supported by a CNPq junior postdoc scholarship.

\end{document}